\documentclass[sigconf]{acmart}
\usepackage{makecell}
\usepackage{graphicx}
\usepackage{booktabs}
\usepackage{multirow}
\usepackage{xcolor} 
\usepackage{threeparttable}
\usepackage{lineno}
\AtBeginDocument{%
  }

\setcopyright{acmlicensed}
\copyrightyear{2018}
\acmYear{2018}
\acmDOI{XXXXXXX.XXXXXXX}
\acmConference[Conference acronym 'XX]{Make sure to enter the correct
  conference title from your rights confirmation emai}{June 03--05,
  2018}{Woodstock, NY}
\acmISBN{978-1-4503-XXXX-X/18/06}

\begin{document}

\title{CliniQ: A Multi-faceted Benchmark for Electronic Health Record Retrieval with Semantic Match Assessment}

\author{Zhengyun Zhao}
\email{zhengyun21@mails.tsinghua.edu.cn}
\affiliation{%
  \institution{Tsinghua University}
  \city{Beijing}
  \country{China}
}

\author{Hongyi Yuan}
\email{yuanhy20@mails.tsinghua.edu.cn}
\affiliation{%
  \institution{Tsinghua University}
  \city{Beijing}
  \country{China}
}

\author{Jingjing Liu}
\email{liuj0513022@163.com}
\affiliation{%
  \institution{Peking Union Medical College}
  \city{Beijing}
  \country{China}}

\author{Haichao Chen}
\affiliation{%
  \institution{Tsinghua Medicine}
  \city{Beijing}
  \country{China}
}
\email{chc16@mails.tsinghua.edu.cn}

\author{Huaiyuan Ying}
\email{yinghy22@mails.tsinghua.edu.cn}
\affiliation{%
  \institution{Tsinghua University}
  \city{Beijing}
  \country{China}
}

\author{Songchi Zhou}
\email{zhou-sc23@mails.tsinghua.edu.cn}
\affiliation{%
 \institution{Tsinghua University}
 \city{Beijing}
 \country{China}}

\author{Yue Zhong}
\email{zhongyue21@mails.tsinghua.edu.cn}
\affiliation{%
 \institution{Tsinghua University}
 \city{Beijing}
 \country{China}}
 
\author{Sheng Yu}
\authornote{Corresponding Author}
\email{syu@tsinghua.edu.cn}
\affiliation{%
 \institution{Tsinghua University}
 \city{Beijing}
 \country{China}}


\begin{abstract}
Electronic Health Record (EHR) retrieval plays a pivotal role in various clinical tasks, but its development has been severely impeded by the lack of publicly available benchmarks. 
In this paper, we introduce a novel public EHR retrieval benchmark, CliniQ, to address this gap.
We consider two retrieval settings: Single-Patient Retrieval and Multi-Patient Retrieval, reflecting various real-world scenarios. 
Single-Patient Retrieval focuses on finding relevant parts within a patient note, while Multi-Patient Retrieval involves retrieving EHRs from multiple patients.
We build our benchmark upon $1,000$ discharge summary notes along with the ICD codes and prescription labels from MIMIC-III, and collect $1,246$ unique queries with $77,206$ relevance judgments by further leveraging powerful LLMs as annotators. 
Additionally, we include a novel assessment of the semantic gap issue in EHR retrieval by categorizing matching types into string match and four types of semantic matches. 
On our proposed benchmark, we conduct a comprehensive evaluation of various retrieval methods, ranging from conventional exact match to popular dense retrievers.
Our experiments find that BM25 sets a strong baseline and performs competitively to the dense retrievers, and general domain dense retrievers surprisingly outperform those designed for the medical domain.
In-depth analyses on various matching types reveal the strengths and drawbacks of different methods, enlightening the potential for targeted improvement.
We believe that our benchmark will stimulate the research communities to advance EHR retrieval systems.
\end{abstract}



\received{20 February 2007}
\received[revised]{12 March 2009}
\received[accepted]{5 June 2009}

\maketitle

\section{INTRODUCTION}
Electronic Health Records (EHRs) are invaluable resources due to the rich patient information they contain \cite{Osmani17, zhang2019high}.
In clinical practice, retrieval is generally the first step to access the information in EHRs: 
physicians need to locate certain information for making clinical decisions \cite{ye2021leveraging, Pampari2018emrQAAL}, and researchers need to search for specific criteria to find patients of interest \cite{li2021patient, bouzille2018drug}. 
Such a process can be time-consuming \cite{arndt2017tethered, ying2025geniegenerativenoteinformation}, and medical practitioners rely heavily on automatic EHR retrieval systems \cite{Hanauer2015SupportingIR, Jackson2017CogStackE}.

\begin{figure*}[tbp]
  \centering
  \includegraphics[width=0.95\textwidth]{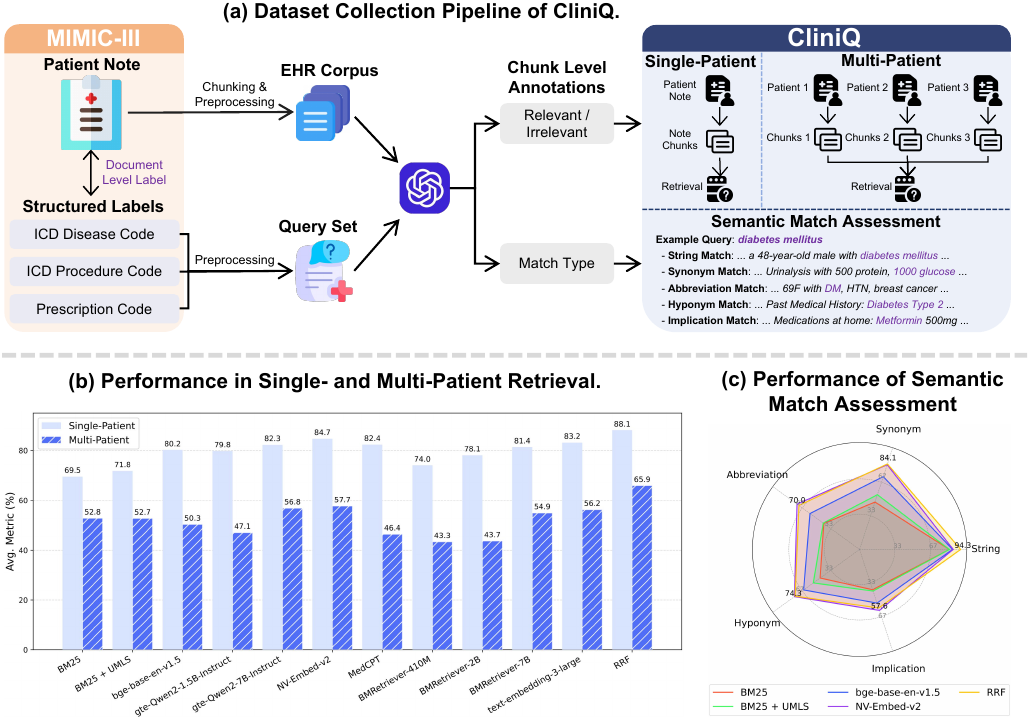}
  \caption{(a) Dataset Collection Pipeline of CliniQ.
  (b) Performance of various baseline retrieval methods in the Single-Patient and Multi-Patient Retrieval settings in CliniQ. The score reported for each model is an average of MRR, NDCG, and MAP for Single-Patient retrieval, and an average of MRR, NDCG@10, and recall@100 for Multi-Patient Retrieval.
  (c) Performance of various baseline retrieval methods regarding different match types in Single-Patient Retrieval. The score reported for each model is an average of MRR, NDCG, and MAP.}
  \label{fig:teaser}
\end{figure*}

However, progress in this field has been limited, and
one of the main reasons is the absence of publicly available benchmarks for evaluating the performance of the EHR retrieval systems. 
Since EHRs contain private patient health conditions, EHR data cannot be distributed or published without deidentification and reviews \cite{Yoon2023EHRSafeGH, pmcp, Keshta2020SecurityAP}. 
Previous research on EHR retrieval relied exclusively on proprietary data \cite{martinez2014improving, Soni2020PatientCR, gupta2024oncoretrievergenerativeclassifierretrieval}.

Therefore, to build a public benchmark for EHR retrieval, a practical approach is to leverage publicly available EHR corpus such as the MIMIC datasets \cite{Johnson2016MIMICIIIAF, johnson2023mimic}.
However, to the best of our knowledge, the only attempt in this direction \cite{Myers2024LessonsLO} did not release their annotated dataset.
The research community still have no accessible benchmark datasets.

Except for accessible corpus, building an EHR retrieval benchmark faces the following main challenges.
Firstly, the query curation and annotation process used to depend entirely on manual experts \cite{Wang2019TestCF, Yang2021ImprovingCE, gupta2024oncoretrievergenerativeclassifierretrieval}, significantly constraining the dataset scale.
The once most widely used dataset, TREC Medical Record tracks \cite{trecmedicalrecord}, which is inaccessible now, incorporates only $85$ queries and $5,895$ positive relevance annotations.
The limited dataset scale might compromise the robustness and generalizability of the benchmark.
Recently, Large Language Models (LLMs) have demonstrated human-level proficiency across a wide array of tasks including relevance judgment \cite{Upadhyay2024ALS, Hosseini2024RetrieveAE}, unlocking new possibilities for this challenge.
However, no studies so far have explored such application of LLMs in the context of evaluating EHR retrieval systems. 

Secondly, existing evaluations generally focuses on one specific downstream application, lacking generalizability in reflecting real-world scenarios \cite{sivarajkumar2024clinical}. 
Broadly speaking, EHR retrieval can be classified into two settings:
\begin{itemize}
    \item \textbf{Single-Patient Retrieval}: identifying relevant parts within one patient's medical records, used in tasks like question answering (QA) \cite{Pampari2018emrQAAL, Lanz2024ParagraphRF} and patient chart review \cite{ye2018extracting, ye2021leveraging}.
    \item \textbf{Multi-Patient Retrieval}: searching for suitable patients in EHR database, used in tasks like patient cohort selection \cite{li2021patient, martinez2014improving} and disease prevalence prediction \cite{bouzille2018drug, hammond2013use}.
\end{itemize}
Concretely, different downstream tasks emphasize different types of queries, ranging from simple terms \cite{ye2018extracting, Ping2021ResearchOS}, natural language questions \cite{Lanz2024ParagraphRF}, to complex criteria \cite{Wang2019TestCF, trecmedicalrecord}.
Despite varied format and complexity, entity retrieval may be seen as an atomic task.
Short terms and questions focusing on certain entity compose a large proportion of real-world EHR queries \cite{Yang2011QueryLA, Natarajan2010AnAO}, and complex criteria with multiple entities and/or logical conditions (such as negative detection) are typically decomposed into single entity queries during retrieval \cite{Jin2021AlibabaDA, Jin2023MatchingPT}.

Thirdly, the semantic gap issue has been a major challenge for the EHR retrieval community \cite{hopkins, tamine2021semantic}.
Specifically, traditional EHR retrieval systems encounter several obstacles \cite{koopman2016information, edinger2012barriers}:
\begin{itemize}
    \item \textbf{Vocabulary Mismatch}: missing synonyms, including abbreviations, of the query. For example, a record containing "RA" may be missed for the query "rheumatoid arthritis".
    \item \textbf{Granularity Mismatch}: missing hyponyms of the query. For example, a record containing "acute tubular necrosis" (a subtype of renal failure) may be missed for the query "acute renal failure".
    \item \textbf{Implication Mismatch}: missing information highly indicative of the query. For example, a record containing "amlodipine" (a common antihypertensive drug) may be missed for the query "hypertension".
\end{itemize}
Despite the significance of quantitatively analyzing the semantic gap issue, there is an absence of sophisticated evaluation frameworks that are capable of revealing the nuanced performance differences on various matching types.
Therefore, the community lacks a clear understanding of the semantic matching abilities of various models.

In this paper, we aim to address these challenges and fill in the blank of a public EHR retrieval benchmark with a novel dataset, CliniQ.
It includes large-scale queries, high-quality annotations, two retrieval settings representing various applications, and categorized labels for semantic match assessment.
The dataset collection pipeline is shown in Figure \ref{fig:teaser} (a).
To be specific, we focus on the task of entity retrieval and include both the settings of Single-Patient Retrieval and Multi-Patient Retrieval.
We build our work upon MIMIC-III.
We utilize chunked MIMIC discharge summaries as EHR corpus and leveraged the ICD-9 disease codes, ICD-9 procedure codes, and prescription labels as queries.
With patient level annotations provided in MIMIC structured database, we use GPT-4o to refine the annotations into chunk level.
Meanwhile, we further combine exact match and GPT-4o to classify the matching types into five categories: string, synonym, abbreviation, hyponym, and implication match.
An example of the five match types is shown in Figure \ref{fig:teaser} (a).
Human evaluation on a subset of CliniQ annotations indicates that GPT-4o highly aligns with medical experts.
We randomly select $1,000$ MIMIC-III discharge summaries, which are split into $16,550$ chunks, as our corpus.
We collect $1,246$ unique queries and $77,206$ detailed relevance judgments, which is an order of magnitude larger than previous datasets.

Based on CliniQ, we comprehensively benchmark various retrievers' performance on EHR retrieval task, including BM25 \cite{bm25}, state-of-the-art dense retrievers of various sizes covering both general and medical domain, and the most capable proprietary embedding model by OpenAI.
An overview of our benchmark results is provided in Figure \ref{fig:teaser} (b) and (c).
In our experiments, the two settings present occasionally different model rankings.
BM25 establishes a quite strong baseline, and the performance is further enhanced in Single-Patient Retrieval and recall@100 in Multi-Patient Retrieval through query expansion.
Dense retrievers show consistent improvement with the increasing parameter size, and general domain retrievers outperform medical domain ones.
With relevance annotations dissected by matching types, we first shed light on the semantic matching abilities of various retrievers in the context of EHR retrieval.
We quantitatively reveal that the advantages of dense retrievers are mainly contributed by semantic matches, and among different types of semantic matches, implication match poses the greatest challenge for retrieval systems.
Moreover, dense retrievers struggle in drug retrieval, where most queries are single-word and annotated by string match.

We believe this benchmark will be a valuable resource for the community, and our comprehensive analysis may point out future research directions.
The benchmark is publicly available through Github (\url{https://github.com/zhao-zy15/CliniQ}) and HuggingFace (\url{https://huggingface.co/datasets/THUMedInfo/CliniQ})\footnote{Due to the credential requirement of MIMIC, we cannot redistribute the corpus directly. Rather, we release the \texttt{hadm\_id}s of patients involved in our benchmark and the script to reproduce the corpus.}.

\section{RELATED WORK}
\subsection{EHR Retrieval Benchmarks}
Typically, a retrieval benchmark has three components: the query set, the corpus, and the relevance judgments.
We review existing EHR retrieval benchmarks from these three perspectives.

\subsubsection{Query set}
According to the downstream task of interest, there are various types of query sets used in EHR retrieval, among which the most popular one is patient cohort criteria \cite{trecmedicalrecord,Wang2019TestCF,Thai2024ACRAB}.
For example, TREC released a total of $85$ queries in 2011 and 2012 Medical Record tracks \cite{trecmedicalrecord}; 
\citet{Wang2019TestCF} employed $56$ real criteria from the Mayo Clinic and Oregon Health \& Science University (OHSU); 
\citet{Thai2024ACRAB} curated $113$ queries focusing on $6$ oncology use cases.
Such queries can be quite complex, potentially including multiple medical entities and complicated logical conditions, such as "Patients taking atypical antipsychotics without a diagnosis schizophrenia or bipolar depression".
Though faithfully reflecting the patient cohort selection scenario, these queries lack generalizability in downstream applications of EHR retrieval, and are actually often decomposed to single entities for better retrieval performance \cite{martinez2014improving, li2021patient, Yang2021ImprovingCE}.

Recently, using medical entities as queries has gained more and more research interest due to their versatility and their alignment with physicians' practices \cite{Yuan2020CODERKC, ruppel2020assessment, Yang2011QueryLA}.
\citet{Ping2021ResearchOS} utilized $8$ cardiovascular disease terms, such as "hypertension" and "palpitation", as queries;
\citet{Yang2021ImprovingCE} incorporated $20$ stroke-related concepts including diseases, symptoms, medications, and procedures; \citet{Shi2022ImprovingNM} extracted $26$ disease mentions from radiology reports as the query set. 
However, all existing works used a quite small number of queries, with limited diversity, often focusing on specific information types or even specific diseases. 
The absence of a large-scale and diverse query set may hinder a comprehensive assessment of the retrieval performance.

\subsubsection{EHR corpus}
Almost all existing evaluations on EHR retrieval use proprietary corpus \cite{trecmedicalrecord, Wang2019TestCF, Thai2024ACRAB}.
To the best of our knowledge, \citet{Myers2024LessonsLO} are the only ones attempting to utilize publicly available EHR corpus, MIMIC-III, to establish an EHR retrieval benchmark.
However, the annotated dataset is not released, and a total of only $50$ patients are incorporated in the evaluation, potentially undermining the robustness of the evaluation.
Besides, previous benchmarks generally adopt either Single-Patient \cite{ye2018extracting, ye2021leveraging, gupta2024oncoretrievergenerativeclassifierretrieval} or Multiple-Patient setting \cite{trecmedicalrecord, Wang2019TestCF, Thai2024ACRAB}.
None has assessed the model performance under different retrieval settings.

\subsubsection{Relevance judgment}
Traditionally, relevance judgment can be obtained solely from human experts \cite{trecmedicalrecord, Wang2019TestCF, Yu2022BIOSAA}, which is prohibitive to scale.
In terms of positive labels, TREC annotated a total of $5,895$ relevant pairs in two years, and \citet{Wang2019TestCF} included $5,815$ positive annotations in their benchmark.
Recently, researchers begin to explore automatic annotation methods to overcome the limitations of dataset scale.
\citet{Thai2024ACRAB} integrates Hypercube \cite{Shekhar2023CouplingSR}, a deterministic reasoning engine based on ontology, in the annotation pipeline and manages to achieve a Cohen's Kappa efficient of $1$ with medical experts.
However, the annotation process still relies on manually reading every patient record and extracting clinical facts regarding the queries, which is not generalizable and scalable.
\citet{Myers2024LessonsLO} automatically annotated three specific types of mentions (diagnosis, surgeries, and antibiotics) using regular expressions and the UMLS knowledge graph.
Despite being totally untethered from intensive human labor, the annotation quality is strictly constrained by exact match, which is known to have a low recall \cite{koopman2016information}.
This limitation may particularly underestimate more advanced methods such as dense retrieval \cite{thakur2021beir}.
In addition, none of existing benchmarks has incorporated detailed assessments on semantic match.

\subsection{EHR Retrieval Applications}
A recent survey \cite{sivarajkumar2024clinical} identifies the primary applications of EHR retrieval as patient chart review (36\%), patient cohort selection (29\%), and disease prevalence prediction (21\%). 
Additionally, EHR retrieval is often regarded as a preliminary step for EHR QA.

\subsubsection{Patient chart review}
Patient chart review refers to the process that clinicians go through a patient's notes to find specific information \cite{ye2018extracting}.
For example, \citet{ye2021leveraging} defines three chart review tasks as retrieving all information related to acute myocardial infarction, Crohn’s disease, and diabetes from the patient notes.
This process can be quite time-consuming due to the vast volume of even one patient's notes.
Therefore, the retrieval step is necessary.
Traditional patient chart review systems, such as EMERSE \cite{Hanauer2015SupportingIR}, relies on exact match methods.
Recently, word embeddings have been applied to enhance the retrieval by providing query recommendation and query expansion \cite{ye2018extracting, Sun2021UsingNI}.
Notably, \citet{ye2021leveraging} proposed a medical term embedding model trained on real clinical notes, and showed significant improvement over general domain embeddings for retrieval performance.

\subsubsection{Patient cohort selection}
In patient cohort selection, physicians or researchers aims to select a group of patients satisfying certain criteria for clinical researches or risk identification \citep{sivarajkumar2024clinical}.
This task can go beyond the domain of text retrieval since some criteria are related to structured patient characteristics such age and gender. 
For example, Cohort Retrieval Enhanced by Analysis of Text from Electronic Health Records (CREATE) is a system that performs cohort selection on both structured and unstructured data using the OMOP Common Data Model and Elasticsearch \cite{Liu2020ImplementationOA}. 
Still, the retrieval process relies largely on fixed vocabulary and Named Entitiy Recognition (NER) tools such as cTAKES \cite{Savova2010MayoCT}.
Facilitated by the data provided by TREC Medical Record tracks, the development of this area has been pushed relatively farther.
More advanced methods including Siamese network \cite{Xiao2020PatientTM} and Transformer-based language model \cite{Soni2020PatientCR} has been applied.

\subsubsection{Disease prevalence prediction}
EHR retrieval can be also applied to predict the prevalence of certain conditions in a population.
For example, \citet{hammond2013use} managed to increase the accuracy of identifying veterans with suicide attempts via searching the medical record database. 
Similarly, \citet{bouzille2018drug} leveraged EHR retrieval to identify 41 additional cases of drug-induced anaphylaxis besides the 25 cases already identified. 

\subsubsection{EHR QA}
EHR QA has been an indispensable component in clinical practice, and EHR retrieval has great potential of significantly improving the efficiency and effectiveness of EHR QA systems \cite{Lanz2024ParagraphRF}.
\citet{Lanz2024ParagraphRF} formally defined the task of EHR retrieval in the context of EHR QA, and evaluated the impacts of various retrieval methods on the performance of downstream QA task using emrQA dataset \cite{Pampari2018emrQAAL}.
They also investigated the effects of different chunking strategies and chunking lengths.

\subsection{Dense Retrieval}
With the rapid development of Pre-trained Language Models (PLMs), dense retrieval has become the predominant retrieval method \cite{neelakantan2022text, Li2023TowardsGT, lee2024nv}.
It leverages dense vector representations of queries and documents generated by PLMs to perform efficient and effective similarity match in high-dimensional embedding spaces \cite{karpukhin2020dense}.
Dense retrievers have been shown to acquire zero-shot abilities through contrastive learning on large-scale paired data \cite{ni2021large, ying2024cortex}, and are thus widely adopted in both academic and industrial scenarios.
One notable example is the \texttt{bge} series models \cite{bge_embedding}, which leverage vast amount of unsupervised pair data mined from web corpus, and are trained with a three-stage pipeline: RetroMAE pre-training \cite{Xiao2022RetroMAEPR}, unsupervised contrastive learning, and fine-tuning on manually annotated data.
Among them, \texttt{bge-base-en-v1.5}, a BERT-based \cite{bert} model with $110$M parameters, demonstrates remarkable capacities on MTEB \cite{muennighoff2022mteb}, the most authoritative embedding benchmark. 

Recently, LLMs present unprecedentedly capacities on various language generation tasks, and researchers start to explore their potential as an embedding model by taking the last token embedding for representation \cite{Ma2023FineTuningLF}.
Generally, these embedding models are initialized from powerful LLMs of various sizes (mostly $1.5\sim 7$B), and are equipped with bidirectional attention and instruct-tuning \cite{lee2024nv}.
The training pipeline usually involves contrastive learning with large-scale unsupervised data, high-quality supervised data, and LLM synthetic data \cite{lee2024nv}.
Some well-known examples include the \texttt{gte-Qwen2} series \cite{Li2023TowardsGT} initialized from \texttt{Qwen2} \cite{qwen2}, and \texttt{NV-Embed-v2} initialized from \texttt{Mistral-7B} \cite{Jiang2023Mistral7}.
The latter is currently the best performing open-source embedding model on MTEB.

In biomedical domain, the development of generalizable dense retrievers has been limited by the lack of high-quality paired data \cite{jin2023medcpt}.
Prior to the advent of LLMs, \citet{jin2023medcpt} introduced MedCPT, which was trained on PubMed user logs and thus specialized in biomedical retrieval.
At the time, MedCPT significantly outperformed general domain encoders in a wide range of biomedical retrieval tasks.
Recently, leveraging powerful LLMs like GPT-3.5 and GPT-4 for data synthesizing, \citet{xu2024bmretriever} introduced BMRetriever, a series of LLM-based embedding model ($410$M, $2$B, and $7$B), further improving the performance on these tasks.
However, the effectiveness of these models on the task of EHR retrieval remains unclear.

\section{BENCHMARK CONSTRUCTION}

\subsection{Corpus Pre-processing}
We randomly sample $1,000$ discharge summary notes from MIMIC-III as our initial corpus.
Following previous works \cite{Mullenbach2018ExplainablePO, Yuan2022CodeSD}, we perform basic data cleaning by removing masks in MIMIC, removing abundant white spaces, and lower-casing all tokens.
We further apply chunking to the notes to fit the context window constraints of dense retrievers and optimize the retrieval performance.
Here we adopt the naive paragraph segmentation by simply splitting the notes into fixed length chunks with overlap since more sophisticated chunking strategies did not reveal consistent and significant improvement in previous work \cite{Lanz2024ParagraphRF}.
Formally, given the cleaned notes $\mathcal{N}=\{ N_1,N_2,\dots\}$, we split each discharge summary $N_i$ into chunks $\mathcal{C}_i = \{c_i^1, c_i^2, \dots\}$, where each chunk $c_i^j$ has a length of $100$ words and two adjacent chunks $c_i^j$ and $c_i^{j+1}$ has 10-word overlap.
Our final corpus consists of the union of all the chunks from separate notes, formally $\mathcal{C}=\bigcup_i\mathcal{C}_i$.

\subsection{Query Curation}
In MIMIC-III structured data, each patient visit is assigned with various ICD-9 disease, ICD-9 procedure, and prescription codes.
Each code corresponds to a term in natural language, and to transform them into user queries suitable for the EHR retrieval task, we apply different processing methods for the three types of codes.

\begin{figure}[tbp]
  \centering
  \includegraphics[width=0.9\linewidth]{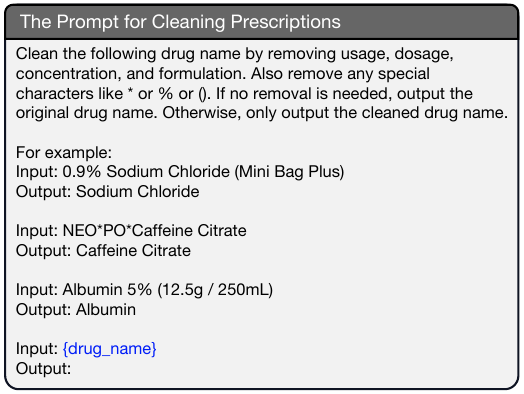}
  \caption{The prompt for drug name cleaning. \textcolor{blue}{drug\_name} represents the input drug name.}
  \label{fig:drug}
\end{figure}

For ICD disease codes, we map all fine-sorted codes to their three-digit ancestors since the classifying system can be so detailed that the disease term becomes far more complicated than typical user queries, or even incomprehensible given the note.
For example, we would map the code "011.92", "\textit{Pulmonary tuberculosis, unspecified, bacteriological or histological examination unknown (at present)}", to code "011", "\textit{Pulmonary tuberculosis}".
Besides, the ICD code terms can contain phrases that are dependent on the coding system, such as "other", "unspecified", and "not elsewhere classified".
To keep the queries self-contained, we manually remove such phrases.
For ICD procedure codes, which are less detailed and complex than the disease codes, we only process the codes by removing these ambiguous terms.
For prescription codes, we utilize the National Drug Code (NDC) provided in MIMIC and remove information on usage, dosage, concentration, and formulation from the drug names using GPT-4o.
The prompt we use is shown in Figure \ref{fig:drug}.

We take the combination of the three types of transformed terms as the query set $\mathcal{Q}_i=\{q_i^1, q_i^2,\dots\}$ for discharge summary $N_i$ of this patient visit.
Since the coding system is universal across patients, we are able to create a global query set by taking union of the query set of all patients in our dataset, formally $\mathcal{Q}=\bigcup_i\mathcal{Q}_i$.

\subsection{Relevance Judgment}
\begin{figure}[htbp]
  \centering
  \includegraphics[width=\linewidth]{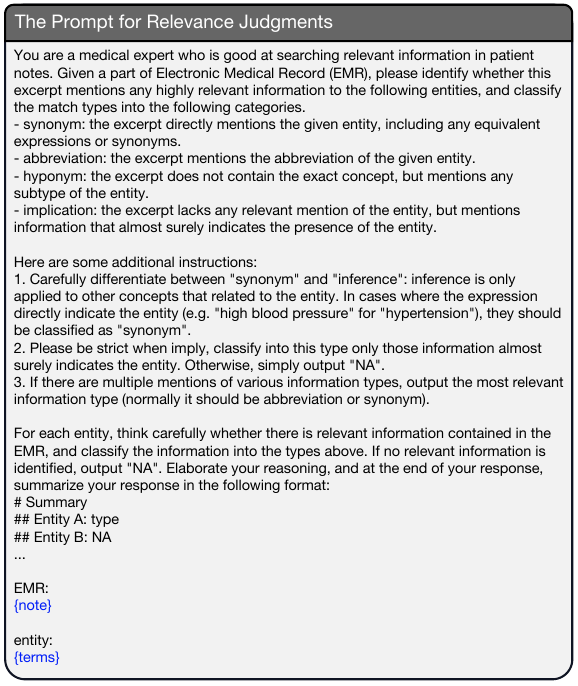}
  \caption{The prompt for relevance judgment.
  \textcolor{blue}{note} and \textcolor{blue}{terms} represents the chunk and the query terms to be annotated.}
  \label{fig:anno}
\end{figure}

The relevance judgment is given at chunk level, and  we first perform a global exact match search for every query to annotate all string matches.
As for semantic matches, we only annotate queries assigned to the patient in the coding systems $\mathcal{Q}_i$, rather than all queries $\mathcal{Q}$, for each note $N_i$ due to cost constraints.
To be specific, for each chunk $c_i^j$ in $N_i$, we prompt GPT-4o\footnote{We strictly conform to the data usage agreement of MIMIC by using Azure OpenAI service with proper certification.} to generate relevance judgment for each query $q_i^k\in \mathcal{Q}_i$ (excluding string matched ones), and further classify the matching type into synonym, abbreviation, hyponym, and implication match.
The formal definitions of the match types can be found in the prompt shown in Figure \ref{fig:anno}.
Formally, GPT-4o takes $c_i^j$ and $\mathcal{Q}_i$ as input, and output a list of judgments $\mathcal{L}_{ij}=\{l_{ij}^1, l_{ij}^2,\dots\}$.
Each $l_i^k$ takes value in $\mathcal{R}=\{\textit{irrelevant}, \textit{synonym}, \textit{abbreviation},\textit{hyponym},\textit{implication}\}$, corresponding to the matching type (or irrelevance) of $q_i^k$.
To enhance the annotation quality, we use Chain-of-Thought (CoT) prompt \cite{Wei2022ChainOT} for relevance judgment.
Combining the two steps above, for each patient note $N_i$, we obtain complete pairwise relevance annotations between queries $\mathcal{Q}_i$ and chunks $\mathcal{C}_i$, classified into five match types (string match and four types of semantic matches).

To evaluate the quality of the automatic annotations, we randomly sample a subset of CliniQ and invite two senior M.D. candidates to annotate the relevance judgments and matching types. 
A chunk record is considered relevant if it contains information that is semantically related to the query and is of interest from the physician's perspective when searching for the query.
Disagreements are discussed to reach a ground truth label.

\section{EXPERIMENTS}
\subsection{Experiment Settings}
\label{sec:setting}
CliniQ contains two retrieval settings.
They share the same corpus, queries, and annotations, but differ in the experiment settings.

In the Single-Patient Retrieval setting, we only consider one patient note $N_i$ during each search, simulating scenarios of individual-level health care.
We observe that not all queries in $\mathcal{Q}_i$ has positive relevance judgments within $N_i$, since some codes may be incomprehensible using merely the discharge summary.
Therefore, for each note $N_i$, we use the terms in $\mathcal{Q}_i$ with at least one relevant chunk within $N_i$ as queries.
The models are required to rank all chunks $\mathcal{C}_i$ in the note so that the relevant chunks are ranked higher.

In the Multi-Patient Retrieval setting, we use chunks of all patient notes included in our dataset $\mathcal{C}$ as corpus, and the union of all terms $\mathcal{Q}$ as queries.
Given each query $q\in \mathcal{Q}$, the model is expected to retrieve all relevant chunks from the whole corpus.

To assess the semantic matching capacities of different models, we additionally provide a benchmark dissected by matching types.
When focusing on certain match type, it is inappropriate to simply treat other relevant pairs as negative, thus we temporarily remove them from the corpus.
For example, in the Single-Patient Retrieval, when we calculate the metrics for synonym matches of query $q_i^k$, we only retain chunks with synonym matches or are irrelevant to $q_i^k$ in the corpus:
$$
\mathcal{C}_i^{\textit{synonym}}=\left\{c_i^j \mid \forall j, l_{ij}^k \in \{\textit{synonym},\textit{irrelevant}\}\right\}
$$
We do not offer a dissected benchmark in the Multi-Patient Retrieval setting due to the potential impacts of false negatives:
the ICD and prescription codes in MIMIC are far from complete \cite{vanAken2021ClinicalOP}, and thus some relevant codes may be missed in our annotation pipeline.
Failing to remove all relevant chunks of other matching types may introduce significant bias to the measurements.
On the other hand, in Single-Patient Retrieval, all pairs of each query and each chunk are annotated, including the irrelevant ones, so the benchmark can faithfully reflect the performance on each matching type.

\subsection{Baselines}
We implement both sparse retrieval with knowledge-graph based query expansion and various state-of-the-art dense retrievers as the baseline models.
Besides, we also include Reciprocal Rank Fusion (RRF) method, combining both lexical and semantical retrievers.

\subsubsection{Sparse retriever}
We implement the Okapi BM25 \cite{bm25} algorithm with default hyperparameter $k_1=1.5$ and $b=0.75$. 
In addition, given the dominant usage of knowledge graph based Query Expansion (QE) in EHR retrieval, we implement a naive QE method with UMLS \cite{Bodenreider2004TheUM}, the most widely used biomedical knowledge graph.
For each query term, we first look it up in UMLS and if a match is found, we expand the query term with its synonyms and hyponyms (\textit{reverse\_is\_a} relationship in UMLS).
We try expanding the query with entities related to the original term via other relationships in UMLS, such as \textit{may\_treat}, to enhance implication match.
It gives suboptimal performance in our experiments, perhaps due to too much noise included.

\subsubsection{Dense retriever}
We include various open-source dense retrievers, covering both general domain retrievers and those specifically designed for the biomedical domain.
For each type, we select models of different parameters and dimension sizes to reveal the effectiveness of scaling in EHR retrievers.
We also include a proprietary embedding from OpenAI.
To be specific, we include the following dense retrievers as our baselines:
\paragraph{Open-source general domain retriever}
\begin{itemize}
    \item \texttt{bge-base-en-v1.5} 
    \item \texttt{gte-Qwen2-1.5B-Instruct} 
    \item \texttt{gte-Qwen2-7B-Instruct} 
    \item \texttt{NV-Embed-v2} 
\end{itemize}

\paragraph{Open-source biomedical domain retriever}
\begin{itemize}
    \item \texttt{MedCPT} 
    \item \texttt{BMRetriever-410M} 
    \item \texttt{BMRetriever-2B} 
    \item \texttt{BMRetriever-7B} 
\end{itemize}

\paragraph{Proprietary retriever}
\begin{itemize}
    \item \texttt{text-embedding-3-large}: the most powerful embedding model by OpenAI.
\end{itemize}

\subsubsection{RRF}
RRF is a simple yet effective algorithm to combine the results from multiple retrievers to yield better performance \cite{Cormack2009ReciprocalRF}. 
For each document, the RRF score for merging $n$ retrievers' results is calculated as $\sum_{i=1}^n\frac{1}{k+r_i}$, where $r_i$ is the rank of the document given by the $i$th retriever, and $k$ is hyperparameter.
In our experiment, we set $k=60$ by convention.
We combine sparse retriever augmented by query expansion and two best-performing dense retrievers, \texttt{NV-Embed-v2} and \texttt{text-embedding-3-large}.
In our experiments, the benefit of incorporating more retrievers is marginal, if any.

\subsection{Evaluation Metrics}
In the Single-Patient Retrieval setting, we evaluate the models with Mean Reciprocal Rank (MRR), Normalized Discounted Cumulative Gain (NDCG), and Mean Average Precision (MAP).
The metrics are not cutoff when computing due to the relatively small corpus.
We also use these metrics for the semantic match assessment under Single-Patient Retrieval. 
In the Multi-Patient Retrieval setting, we utilize MRR and NDCG@10 as shallow metrics to measure the accuracy of the top-ranked results, and employ recall@100 as a deep metric to assess the recall capability of the models.

\section{RESULTS}

\subsection{Dataset Statistics}
The basic statistics of CliniQ are shown in Table \ref{tab:stat}.
We report the total number of queries, chunks, and relevance judgments for Multi-Patient Retrieval, and report the average number per patient record for Single-Patient Retrieval.
Starting from 1,000 discharge summaries in MIMIC-III, we split them into 16,550 chunks of 100 words, with an average of 16.6 chunks per document (Q1: 11.0; Q3: 21.0).
The longest record in our corpus consists of 55 chunks.
We collect 1,246 queries in total, comprising 329 diseases, 365 clinical procedures, and 552 drugs.
The collected queries cover 50.2\% of the three-digit ICD-9 disease codes (329/655) and 13.6\% of the ICD-9 procedure codes (365/2679).
The coverage of drug codes is inaccessible due to the lack of a standardized drug vocabulary\footnote{The NDC system is not a well-controlled dictionary like ICD. It lacks a one-to-one mapping between drug concepts and unique identifiers.}.
In Single-Patient Retrieval, each record is associated with 27.9 queries on average (Q1: 18; Q3: 37; Max: 84), which results in a total of nearly 28k searches in this setting (within each record, we perform one search for each query associated).

\begin{table}[tbp]
  \centering
  \begin{threeparttable}
  \caption{Dataset Statistics of CliniQ.}
  \label{tab:stat}
  \begin{tabular}{lcc}
    \toprule
    & \textbf{Single-Patient\tnote{*}} & \textbf{Multi-Patient}\\
    \midrule
    Patient Record & - & 1,000\\
    Chunk of Records & 16.6 & 16,550\\
    \midrule
    Query & 27.9 & 1,246\\
    \quad - Disease & 6.8 & 329\\
    \quad - Procedure & 2.9 & 365\\
    \quad - Drug & 18.2 & 552\\
    \midrule
    Relevance Judgment\tnote{**} & 77.3 & 77,206\\
    \quad - String & 29.2 & 29,149\\
    \quad - Synonym & 24.8 & 24,798\\
    \quad - Abbreviation & 3.0 & 3,039\\
    \quad - Hyponym & 4.3 & 4,288\\
    \quad - Implication & 15.9 & 15,932\\
    \bottomrule
  \end{tabular}
  \begin{tablenotes}  
    \footnotesize  
    \item[*] In Single-Patient Retrieval, we report the average numbers per patient.
    \item[**] We only count positive labels, i.e. the number of relevant pairs.
  \end{tablenotes}  
  \end{threeparttable}
\end{table}

\begin{figure}[tbp]
  \centering
  \includegraphics[width=\linewidth]{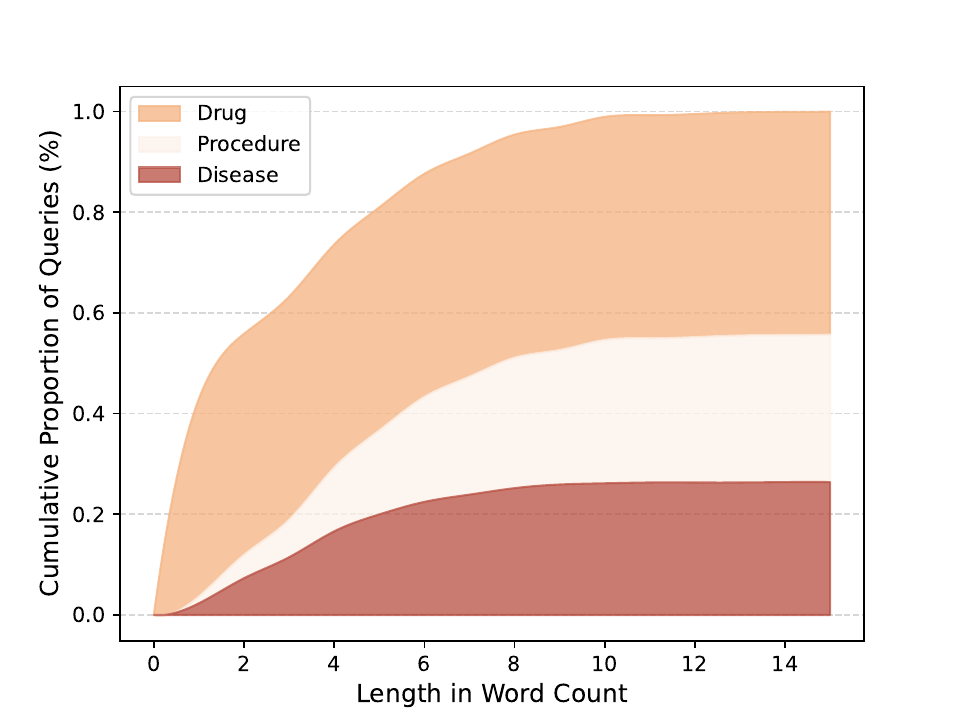}
  \caption{Cumulated Proportion of query length in word counts.}
  \label{fig:query_len}
\end{figure}

The distribution of the query length in word count is shown in Figure \ref{fig:query_len}.
Notably, 43\% of the queries are single-word and mostly drug names.
All drug queries are less than 4 words, while the length distribution of disease and procedure queries are more even.
On average, a query in CliniQ has 3.14 words (Q1: 1.0; Q3: 5.0; Max: 15).

\begin{figure}[tbp]
  \centering
  \includegraphics[width=\linewidth]{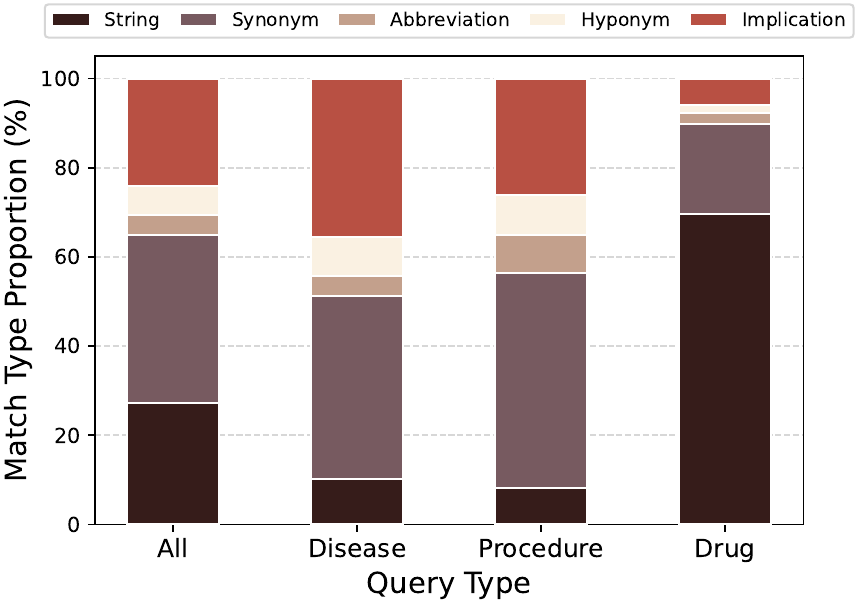}
  \caption{Distributions of different match types decomposed by the query type.}
  \label{fig:query_sem}
\end{figure}

Combining string match and GPT-4o annotations, we collect over 77k pairs of relevance judgments with fine-grained match types.
Among all pairs, over 29k (about 38\%) are exact string matches, and among the semantic matches, most are synonym matches (25k, 32\%) and implication matches (16k, 21\%).
Decomposing the query set according to different query types, we observe a notable variation in distributions of different match types, as shown in Figure \ref{fig:query_sem}.
Relevance annotations related to drug queries contain nearly 70\% string match, while the corresponding proportion in disease and procedure queries are both less than 10\%.
Abbreviation, hyponym, and implication matches combined comprise less than 10\% of relevance annotations.

Human evaluation is conducted on a sample of 141 chunks with 5,221 automatic annotations, derived from 10 randomly chosen patient notes.
The subset for human evaluation covers 181 unique queries (about 10\% of all queries in CliniQ) and is thus representative of the overall dataset quality.
Table \ref{tab:human} presents the inter-annotator agreement between GPT-4o's outputs, two human annotators, and the ground truth. 
The evaluation demonstrates that our automatic annotations achieve strong concordance with the ground truth, performing comparably to medical experts in both relevance judgment and matching type classification. 
The primary errors observed in both automated and manual annotations stem from over-implication: specifically, the assignment of \textit{implication} labels to tenuously relevant cases.

\begin{table}[tbp]
  \centering
  \begin{threeparttable}
  \caption{Human Evaluation Results. For relevance annotations, we only consider binary labels (relevant v.s. irrelevant) and report the Cohen's Kappa coefficient.
  For match types, we formulate the annotation as a multi-class classification and use accuracy metric.}
  \label{tab:human}
  \begin{tabular}{lccc}
    \toprule
    & \textbf{GPT-4o} & \textbf{Annotator 1} & \textbf{Annotator 2}\\
    \midrule
    Relevance & 0.985 & 0.989 & 0.992\\
    Match Types & 0.995 & 0.998 & 0.996\\
    \bottomrule
  \end{tabular}
  \end{threeparttable}
\end{table}

\begin{table*}[tbp]
  \centering
  \begin{threeparttable}
  \caption{Performance of various retrieval methods on CliniQ.}
  \label{tab:overall}
  \begin{tabular}{lcccccccc}
    \toprule
    \multirow{2}{*}[-0.8ex]{\textbf{Model}} & \multirow{2}{*}[-0.8ex]{\textbf{Size}} & \multirow{2}{*}[-0.8ex]{\textbf{Dimension}} & \multicolumn{3}{c}{\textbf{Single-Patient}} & \multicolumn{3}{c}{\textbf{Multi-Patient}}\\
    \cmidrule(r){4-6} \cmidrule(r){7-9}
    & & & \textbf{MRR} & \textbf{NDCG} & \textbf{MAP} & \textbf{MRR} & \textbf{NDCG@10} & \textbf{Recall@100}\\
    \midrule
    BM25 & - & - & 71.65 & 74.52 & 62.42 & 59.18 & 60.26 & 39.01\\
    \quad+ UMLS & - & - & 74.14 & 76.34 & 64.91 & 57.81 & 59.87 & 40.53\\
    \midrule
    \texttt{bge-base-en-v1.5} & 110M & 768 & 82.48 & 83.59 & 74.54 & 54.97 & 56.51 & 39.50\\
    \texttt{gte-Qwen2-1.5B-Instruct} & 1.5B & 1536 & 81.94 & 83.28 & 74.16 & 50.70 & 52.43 & 38.03\\
    \texttt{gte-Qwen2-7B-Instruct} & 7B & 3584 & 84.59 & 85.33 & 77.02 & \underline{60.39} & \underline{62.06} & 48.04\\
    \texttt{NV-Embed-v2} & 7B & 4096 & \underline{86.57} & \underline{87.36} & \underline{80.21} & 59.48 & \underline{62.06} & \underline{51.54}\\
    \midrule
    \texttt{MedCPT} & 220M\tnote{*} & 768 & 84.23 & 85.49 & 77.42 & 47.21 & 50.07 & 41.97\\
    \texttt{BMRetriever-410M} & 410M & 1024 & 76.07 & 78.49 & 67.57 & 48.23 & 50.07 & 31.71\\
    \texttt{BMRetriever-2B} & 2B & 2048 & 80.31 & 81.89 & 72.15 & 46.68 & 49.13 & 35.16\\
    \texttt{BMRetriever-7B} & 7B & 4096 & 83.68 & 84.55 & 75.92 & 58.98 & 60.76 & 45.08\\
    \midrule
    \texttt{text-embedding-3-large} & - & 3072 & 85.16 & 86.09 & 78.36 & 59.54 & 60.45 & 48.75\\
    \midrule
    RRF & - & - & \textbf{89.92} & \textbf{90.18} & \textbf{84.32} & \textbf{67.04} & \textbf{68.74} & \textbf{61.92}\\
    \bottomrule
  \end{tabular}
  \begin{tablenotes}  
    \footnotesize  
    \item[*] \texttt{MedCPT} has separate query encoder and document encoder, so we count the parameter size as the summation of both models.
  \end{tablenotes}  
  \end{threeparttable}
\end{table*}

\begin{table}[tbp]
  \caption{Performance of various retrieval methods on Single-Patient Retrieval, dissected by match types.}
  \label{tab:match}
  \begin{tabular}{lcccc}
    \toprule
    \textbf{Model} & \textbf{Match Type} & \textbf{MRR} & \textbf{NDCG} & \textbf{MAP}\\
    \midrule
    BM25 & String & 83.92 & 86.25 & 81.09\\
    & Synonym & 44.76 & 55.62 & 39.45\\
    & Abbreviation & 38.58 & 50.86 & 34.47\\
    & Hyponym & 42.76 & 54.44 & 39.23\\
    & Implication & 36.30 & 50.22 & 32.25\\
    \midrule
    BM25 + UMLS & String & 83.86 & 86.24 & 81.18\\
    & Synonym & 53.12 & 61.50 & 46.83\\
    & Abbreviation & 38.40 & 51.60 & 35.70\\
    & Hyponym & 51.19 & 61.10 & 47.64\\
    & Implication & 38.13 & 51.19 & 33.43\\
    \midrule
    \texttt{bge-base-en-v1.5} & String & 87.35 & 88.93 & 83.96\\
    & Synonym & 72.48 & 76.45 & 65.78\\
    & Abbreviation & 55.15 & 64.55 & 51.74\\
    & Hyponym & 63.34 & 70.52 & 59.41\\
    & Implication & 51.70 & 61.05 & 45.51\\
    \midrule
    \texttt{NV-Embed-v2} & String & 87.67 & 89.50 & 84.85\\
    & Synonym & 84.29 & 86.17 & 79.37\\
    & Abbreviation & 71.50 & 76.91 & 67.97\\
    & Hyponym & 74.41 & 79.39 & 71.40\\
    & Implication & 59.59 & 67.04 & 53.25\\
    \midrule
    RRF & String & 94.34 & 95.26 & 93.21\\
    & Synonym & 84.97 & 86.82 & 80.42\\
    & Abbreviation & 69.17 & 75.09 & 65.59\\
    & Hyponym & 73.72 & 78.74 & 70.45\\
    & Implication & 56.94 & 65.05 & 50.78\\
    \bottomrule
  \end{tabular}
\end{table}

\begin{table*}[htbp]
  \caption{Performance of various retrieval methods on different types of queries.}
  \label{tab:query}
  \begin{tabular}{lccccccc}
    \toprule
    \multirow{2}{*}[-0.7ex]{\textbf{Model}} & \multirow{2}{*}[-0.6ex]{\textbf{\makecell[c]{Query\\Type}}} & \multicolumn{3}{c}{\textbf{Single-Patient}} & \multicolumn{3}{c}{\textbf{Multi-Patient}}\\
    \cmidrule(r){3-5} \cmidrule(r){6-8}
    & & \textbf{MRR} & \textbf{NDCG} & \textbf{MAP} & \textbf{MRR} & \textbf{NDCG@10} & \textbf{Recall@100}\\
    \midrule
    BM25 & Disease & 68.01 & 71.42 & 54.65 & 39.62 & 41.26 & 20.41\\
    & Procedure & 66.75 & 71.12 & 56.55 & 32.57 & 34.54 & 33.55\\
    & Drug & 73.78 & 76.22 & 66.24 & 88.43 & 88.59 & 53.70\\
    \midrule
    BM25 + UMLS & Disease & 73.81 & 75.13 & 59.55 & 42.96 & 46.24 & 24.09\\
    & Procedure & 70.78 & 74.04 & 60.48 & 34.76 & 37.36 & 35.15\\
    & Drug & 74.80 & 77.16 & 67.61 & 81.90 & 82.87 & 53.88\\
    \midrule
    \texttt{bge-base-en-v1.5} & Disease & 80.05 & 80.49 & 67.41 & 44.23 & 47.35 & 29.80\\
    & Procedure & 78.71 & 80.12 & 68.66 & 38.15 & 41.49 & 44.81\\
    & Drug & 83.99 & 85.29 & 78.12 & 72.49 & 71.91 & 41.77\\
    \midrule
    \texttt{NV-Embed-v2} & Disease & 85.83 & 85.28 & 74.75 & 55.16 & 58.32 & 41.01\\
    & Procedure & 85.20 & 85.88 & 77.38 & 48.70 & 52.78 & 60.86\\
    & Drug & 87.06 & 88.37 & 82.69 & 69.19 & 70.42 & 51.66\\
    \midrule
    RRF & Disease & 87.82 & 86.64 & 76.52 & 55.06 & 58.61 & 45.62\\
    & Procedure & 85.48 & 86.24 & 77.93 & 47.69 & 52.01 & 61.84\\
    & Drug & 91.40 & 92.12 & 88.22 & 86.98 & 85.85 & 71.70\\
    \bottomrule
  \end{tabular}
\end{table*}

\subsection{Benchmark Results}
The performance of various baseline retrieval methods are shown in Table \ref{tab:overall}.
We observe that the performance rankings of the models are inconsistent across the two retrieval settings. 
For example, while \texttt{bge} model presented much higher performance over BM25 in Single-Patient Retrieval, its performance in Multi-Patient Retrieval is comparable to, or even lower than BM25.
This discrepancy suggests that these two scenarios emphasize different kinds of abilities, highlighting the need for a comprehensive evaluation on both settings.

Though performing poorly in Single-Patient Retrieval, BM25 establishes a quite strong baseline in Multi-Patient Retrieval, surpassing all small-scale (<7B) dense retrievers in terms of shallow metrics.
Query expansion based on UMLS knowledge graph significantly enhance the performance of sparse retrieval in Single-Patient Retrieval, but still lags behind dense retrievers a lot.
In Multi-Patient Retrieval, query expansion brings a bit benefits to recall but causes a slightly lower MRR and NDCG, which may be attributed to the noisy nature of knowledge graph.

Among the dense retrievers, we observe a consistent improvements with growing parameter size and embedding dimension, demonstrated by the \texttt{gte-Qwen2} and \texttt{BMRetriever} series models.
\texttt{NV-Embed-v2}, the currently top 1 open-source embedding model on MTEB, achieves the best results across both settings, even surpassing proprietary embedding model by OpenAI.
Among dense retrivers with a parameter size less than 7B, \texttt{MedCPT} presents the best results in Single-Patient Retrieval.
In Multi-Patient Retrieval, \texttt{MedCPT} also shows superior performance in terms of recall@100, while \texttt{bge-base-en-v1.5} gives the best results in shallow metrics.
Generally speaking, dense retrievers specifically designed for and trained in biomedical domain perform suboptimally compared to general domain retrievers.
This is likely due to the fact that EHRs still lie out of the distribution of the training corpus used by traditional biomedical models.
This significant discrepancy highlights the need for future efforts on retrieval methods tailored for the task of EHR retrieval.

RRF brings in huge improvements over all baseline methods, especially in Multi-Patient Retrieval.
Compared to \texttt{NV-Embed-v2}, RRF elevates recall@100 from 51.54 to 61.92, and MRR from 59.48 to 67.04.
The superior performance of RRF underscores the importance of combining lexical and semantic matching abilities in EHR retrieval, which we will investigate deeper in the next section.

\subsection{Semantic Match Assessment}
\label{sec:sem}
With each relevance judgment classified into five categories, we provide a detailed semantic match assessment under the Single-Patient Retrieval setting in Table \ref{tab:match}.
For brevity, we only include sparse retrieval, two dense retrievers, and RRF in the table.
As expected, BM25 presents strong capacities in string match, achieving an MRR of over 80\%, but it struggles in semantic matching.
With UMLS-based query expansion, the semantic matching ability is greatly enhanced, especially in synonym and hyponym match, with an increase of 8\% in MRR each.

The detailed benchmark reveals that the performance differences between sparse and dense retrieval are mainly contributed by semantic matches, so is the advantage of large-scale models over small-scale models.
On the other hand, compared to \texttt{NV-Embed-v2}, RRF is mainly superior on string match, yet the semantic matching ability is compromised a bit except for synonym matches.

Comparison across different types of semantic match shows that implication matches pose the greatest challenges to all baseline retrieval methods, with the highest MRR being 59.59 achieved by \texttt{NV-Embed-v2}.
Besides, there is still remarkable room for improvement regarding abbreviation and hyponym matches.
Though collectively compose less than 10\% of the relevance judgments in CliniQ, the insufficient capacities in retrieving abbreviations and hyponyms call for further research efforts.

\subsection{Query type assessment}
We also conduct detailed analysis regarding different query types: disease, procedure, and drugs.
We consider the same baseline methods as in Section \ref{sec:sem}, and the results are shown in Table \ref{tab:query}.
In Simple-Patient Retrieval, generally speaking, model performance on disease queries and procedure queries are comparable, but significantly lower than drug queries across all methods.
This may be attributed to the unique characteristics of drug queries: most of them are single-word and annotated through exact string match.
These features also account for the fact that BM25 ranks top 1 among all retrievers including RRF in terms of shallow metrics in Multi-Patient Retrieval of drugs.
Specifically, query expansion, despite a slightly higher recall, causes a significant drop in MRR and NDCG@10.
Dense retrievers behave suboptimally in drug retrieval under Multi-Patient setting, indicating that using dense representations to match detailed information such as drug mentions may still be a huge challenge.
RRF, a combination of sparse and dense retrieval results leads to a decreased shallow metrics and a significant higher recall.

As for performance on disease and procedure queries in 
Multi-Patient Retrieval, procedures generally presents a lower MRR and NDCG, but a much higher recall, which can be explained by the much sparser annotations of procedure queries (25 relevant chunk per procedure query v.s. 98 relevant chunk per disease query on average).

\section{FUTURE RESEARCH DIRECTIONS}
Through a comprehensive analysis of the performance of various retrieval methods on the task of EHR retrieval, we point out several potential future research directions.
Firstly, we observe that general domain retrievers outperform medical domain ones, highlighting the significance of the discrepancy between EHRs and conventional biomedical training corpus.
The advantage of general domain retrievers over BM25 on CliniQ is also moderate, calling for researches tailored for the task of EHR retrieval.

Secondly, the inferior performance of dense retrievers on drug retrieval (single-word string match) might indicate their insufficiency in string match, which is also observed in general domain retrieval \cite{Sciavolino2021SimpleEQ, Arabzadeh2021PredictingET, Zhuang2023TyposawareBP}.
The drug mentions in EHR are typically superficial, and can even be buried within extensive medication lists.
Retaining such detailed information in dense representations is a challenging yet highly valuable research area.

Thirdly, we reveal in our experiments that implication matches pose the greatest challenges to all methods.
This type of semantic match requires the retriever to be equipped with extensive medical knowledge and even medical reasoning abilities.
The superiority of \texttt{NV-Embed-v2} may present a potential solution: scaling the model parameter brings in huge benefits in knowledge and reasoning capacities.
Adopting the capabilities of LLMs for EHR retrieval remains an important research topic.

Last but not least, RRF combining sparse and dense retrievers leads to remarkable improvements in CliniQ, yet the semantic matching abilities of dense retrievers are compromised a bit.
Therefore, finding more efficient and effective methods to leverage both lexical and semantic matches may be of great significance.

\section{CONCLUSION}
In this paper, we introduce CliniQ, a public benchmark for EHR retrieval, addressing the need for accessible evaluation resources in this area. 
CliniQ is a magnitude larger than previous benchmarks in terms of the numbers of both queries and relevance judgments.
CliniQ supports both Single-Patient and Multi-Patient Retrieval settings, providing a multi-faceted evaluation.
CliniQ also enables detailed analysis regarding the semantic gap issue.
We conduct comprehensive analysis on CliniQ, and demonstrate that BM25 provides a strong baseline.
In our experiments, general domain dense retrievers outperform those tailored for the medical domain. 
We also highlight the strengths and weaknesses of various methods regarding various match types.
CliniQ aims to advance EHR retrieval research by providing a versatile, robust, and publicly available benchmark, fostering improvements in retrieval systems for better clinical outcomes.

\section*{Limitations}
This study has several limitations that warrant discussion. 
First, our benchmark currently focuses exclusively on discharge summaries, while inclusion of diverse clinical note types would better reflect real-world challenges. 
However, comprehensive annotation of all clinical notes, which vastly outnumber discharge summaries, would be prohibitively expensive for us, with even 1,000 patients potentially requiring tens of thousands of dollars in annotation costs.

Second, although our benchmark covers most commonly searched entities, it omits certain categories such as symptoms and anatomical structures.
This limitation stems from the absence of standardized coding systems (comparable to ICD) for these entities, which form the basis of our query collection. 
Furthermore, without a unified vocabulary system, we cannot effectively cluster identical queries across different patients or note chunks to enhance benchmark construction.

Third, as noted in Section \ref{sec:setting}, the incompleteness of ICD labels introduces unavoidable false negatives into our benchmark. 
While exhaustive annotation of all 1,246 queries against 16,550 chunks in CliniQ would be computationally intractable (a challenge shared by all retrieval benchmarks, actually), established benchmarks have nonetheless proven valuable for fair evaluation and have significantly advanced retrieval research over time.

\bibliographystyle{ACM-Reference-Format}
\bibliography{sample-base}





\end{document}